\documentclass[prl,preprint,amsmath,amssymb]
{revtex4}
\usepackage{graphicx}
\usepackage{dcolumn}
\usepackage{bm}
\newcommand{\beq}{\begin{equation}}
\newcommand{\eeq}{\end{equation}}
\newcommand{\ba}{\begin{array}{ccc}}
\newcommand{\ea}{\end{array}}

\newcommand{\br}{{\bm r}}
\newcommand{\bk}{{\bm k}}

\newcommand{\bq}{{\bm q}}
\newcommand{\bK}{{\bm K}}
\newcommand{\bQ}{{\bm Q}}
\def\bea{\begin{eqnarray}}
\def\eea{\end{eqnarray}}
\usepackage{txfonts}
\usepackage{mathrsfs}
\usepackage{feynmf}
\usepackage{comment}
\usepackage{float}
\preprint{arXiv:1303.2114}

\begin{document}
\title{Bond order in two-dimensional metals\\ with antiferromagnetic exchange interactions} 

\author{Subir Sachdev}
\affiliation{Department of Physics, Harvard University, Cambridge MA 02138}
\author{Rolando La Placa}
\affiliation{Department of Physics, Harvard University, Cambridge MA 02138}

\date{\today\\
\vspace{1.6in}}

\begin{abstract}
We present an unrestricted Hartree-Fock computation of charge-ordering instabilities of two-dimensional
metals with antiferromagnetic exchange interactions, allowing for arbitrary ordering wavevectors and internal wavefunctions
of the particle-hole pair condensate. We find that the ordering has a dominant $d$ symmetry of rotations about
lattice points for a range of ordering wavevectors, including those observed in recent experiments at low temperatures
on YBa$_2$Cu$_3$O$_y$.
This $d$ symmetry implies the charge ordering is primarily on the bonds of the Cu lattice, and
we propose incommensurate bond order parameters for the underdoped cuprates.
The field theory for the onset of N\'eel order in a metal has an emergent pseudospin symmetry 
which `rotates' $d$-wave Cooper
pairs to particle-hole pairs (Metlitski {\em et al.}, Phys. Rev. B  {\bf 82}, 075128 (2010)):
our results show that this symmetry has consequences even when the spin correlations are short-ranged
and incommensurate.
\end{abstract}


\maketitle

A remarkable series of experiments \cite{jenny,ali,kohsaka,eric,lawler,mesaros,julien,kohsaka2,keimer,chang,hawthorn,hawthorn2,proust} 
have shed new light on the underdoped region of the 
cuprate high temperature superconductors. These experiments detect a
bi-directional charge density wave with a period in the range of 3 to 5 lattice spacings at low hole densities and low temperatures ($T$).
This order is co-incident with regions of the phase diagram where quantum oscillations \cite{louis}
were observed in YBa$_2$Cu$_3$O$_y$, strongly supporting the hypothesis \cite{julien,louis2,suchitra1,suchitra2,julien2} that the 
charge density wave is responsible
for the Fermi pockets leading to quantum oscillations. Some of the experiments \cite{kohsaka,lawler,mesaros,julien,kohsaka2,hawthorn,hawthorn2,park,damle}
also indicate that there is negligible modulation
of the charge density on the Cu sites; instead, it is primarily a {\em bond\/} density wave, 
with modulations
in spin-singlet observables on the Cu-Cu links, such as the electron kinetic energy.

This paper presents a Hartree-Fock computation of charge-ordering
instabilities of a two-dimensional metal of electrons with antiferromagnetic exchange interactions
(described by a `$t$-$J$' model). 
We allow the charge-ordering to appear at any wavevector, $\bQ$,
and also allow an arbitrary internal wavefunction, $\Delta_\bQ (\bk)$ 
for the spin-singlet particle-hole pair condensate responsible for the density wave order (here $\bQ$ is the center-of-mass
momentum of the particle-hole pair, and $\bk$ is the relative momentum).
We show that this freedom leads to significant insight, despite the simplicity of our method.
We find that for a range of small $\bQ$ (more precisely, in the `$\mathcal{T}$ preserved' region of
Fig.~\ref{fig:lambda}), including those observed so far in the 
experiments \cite{jenny,ali,kohsaka,eric,lawler,mesaros,kohsaka2,keimer,chang,hawthorn,hawthorn2} at low $T$, 
the dominant structure of the internal
wavefunction has a $d$-wave form \cite{max}, with $\Delta_\bQ (\bk) \sim (\cos k_x - \cos k_y)$ for a band-structure appropriate
for the cuprates. This $d$ symmetry implies that the charge order
is located primarily on the Cu-Cu links, there is little modulation of the charge density on the Cu sites, and
time-reversal symmetry ($\mathcal{T}$) is preserved.
We refer to this class of charge order as an `incommensurate $d$-wave bond order'.
Our computation also allows for other spin-singlet orders, such as Ising-nematic \cite{KFE98,YK00,HM00}, `$d$-density wave'  \cite{kotliar,sudip,leewen}, and `circulating currents' \cite{varma}, the last two of which
break $\mathcal{T}$: they are all less preferred than the incommensurate $d$-wave bond order in the underdoped region,
while Ising-nematic order is preferred at larger doping.

\begin{figure}
\begin{center}
 \includegraphics[width=3.1in]{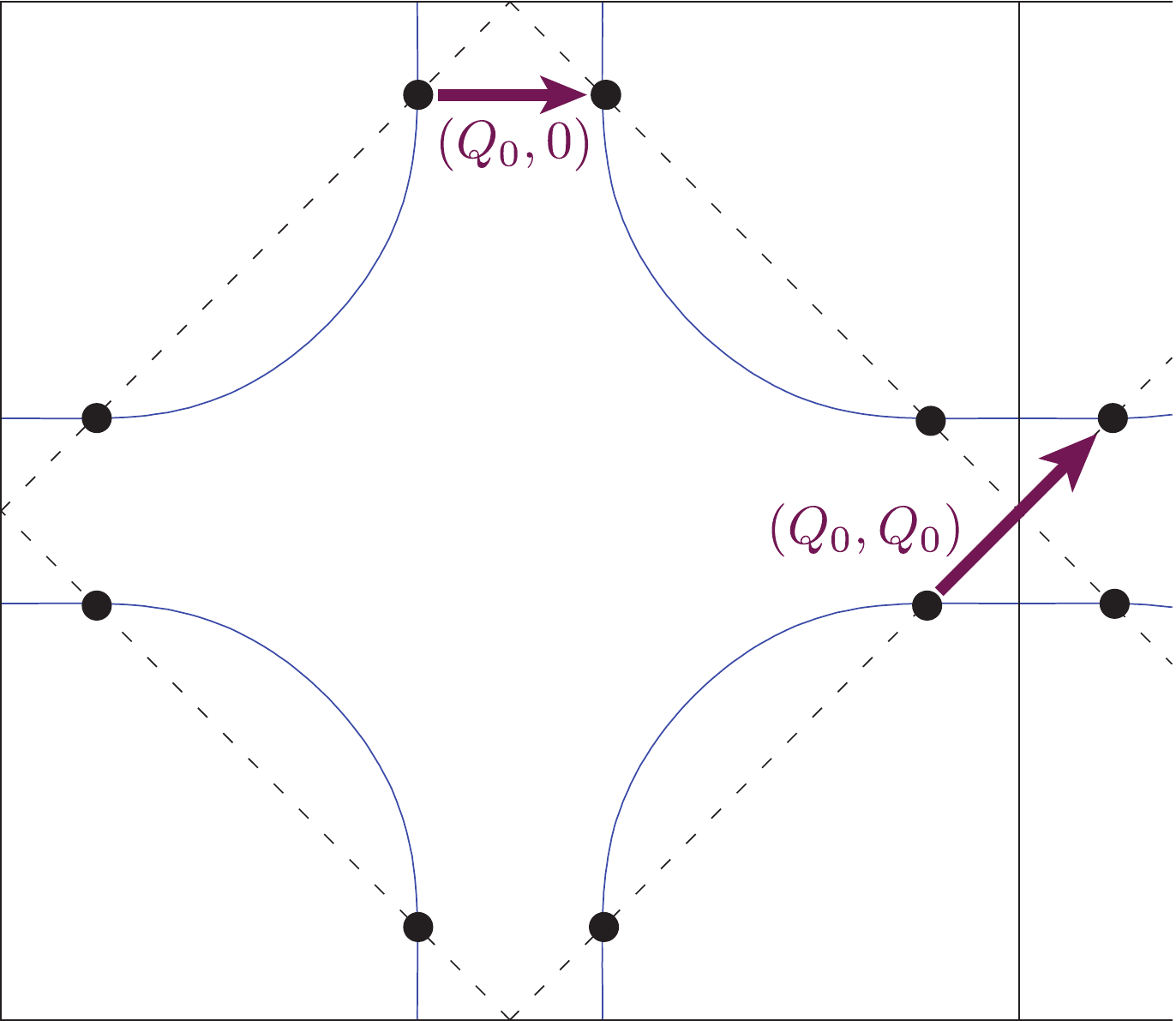}
 \caption{Fermi surface of the hole-doped cuprates, showing the Brillouin zone boundary to antiferromagnetism at $\bK = (\pi,\pi)$
 (dashed lines),
 the hot spots (filled circles), and the wavevectors $(Q_0,0)$ and $(Q_0,Q_0)$.
 }
\label{fig:bz}
\end{center}
\end{figure}

\begin{figure}
\begin{center}
 \includegraphics[width=3.8in]{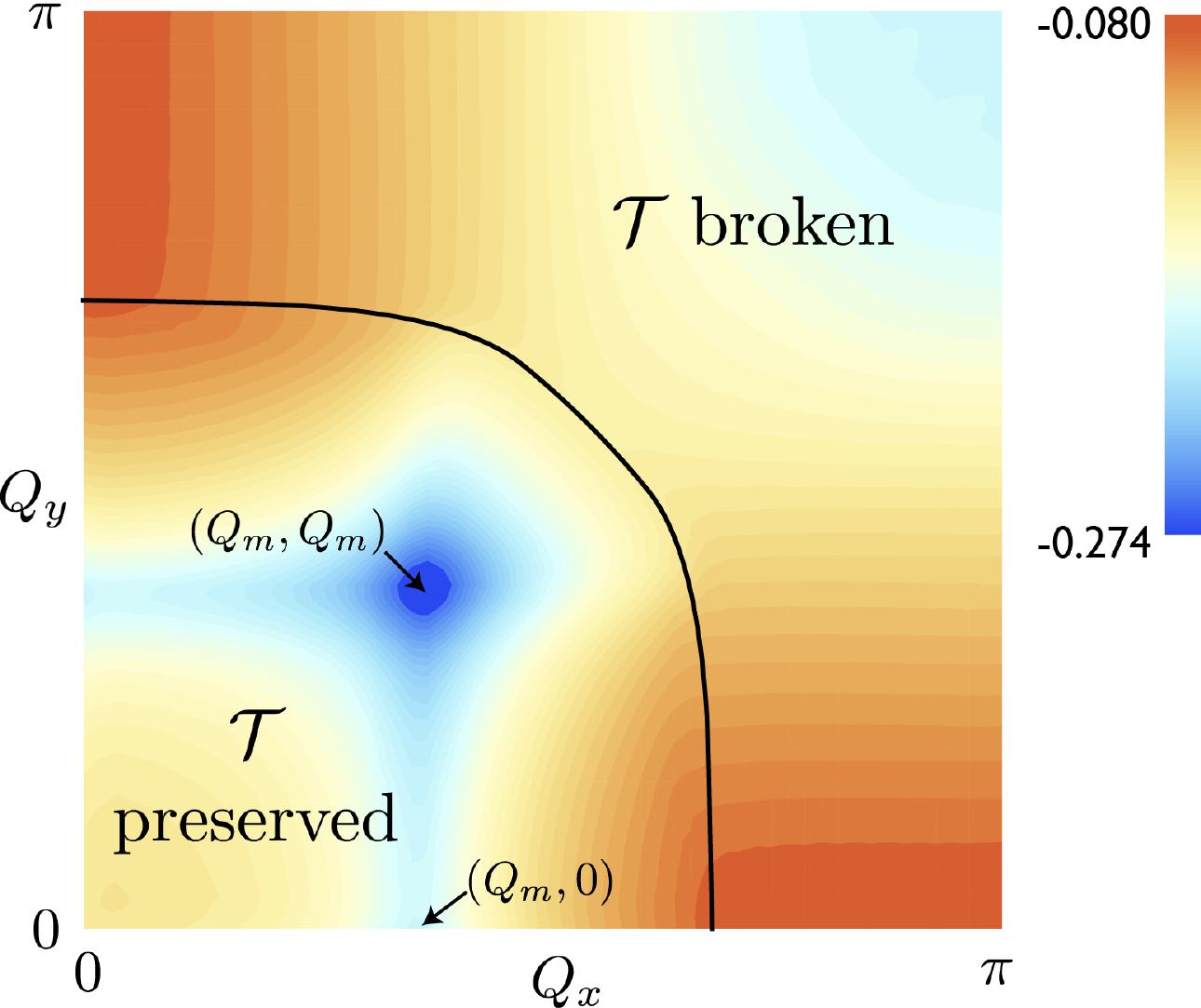}
 \caption{Plot of $\lambda_\bQ/A$, where $\lambda_\bQ$ is the smallest charge order eigenvalue, as a function of $Q_x$ and $Q_y$.
 We used $\mu = -1.11856$, $\xi = 2$,
 $T=0.06$, $\delta=1/4$ and $L=64$.
 Charge order appears when $\lambda_{\bQ} < -1$, which happens when $A$ is large enough. 
 The global minimum is at $(Q_m,Q_m)$ and $Q_m$ is plotted in Fig.~\ref{fig:min} as a function of $\mu$.
 Notice also the blue valleys extending from $(Q_m,Q_m)$ to $(Q_m,0)$ and $(0,Q_m)$. 
 The region with time-reversal, $\mathcal{T}$, preserved has the eigenfunctions
 $\Delta_\bQ (-\bk) = \Delta_\bQ (\bk)$ which are predominantly $d$, 
 while the region with $\mathcal{T}$ broken has  $\Delta_\bQ (-\bk) = -\Delta_\bQ (\bk)$, 
 as shown for some values of $\bQ$ in Table~\ref{table}.
 }
\label{fig:lambda}
\end{center}
\end{figure}

\begin{figure}
\begin{center}
 \includegraphics[width=3in]{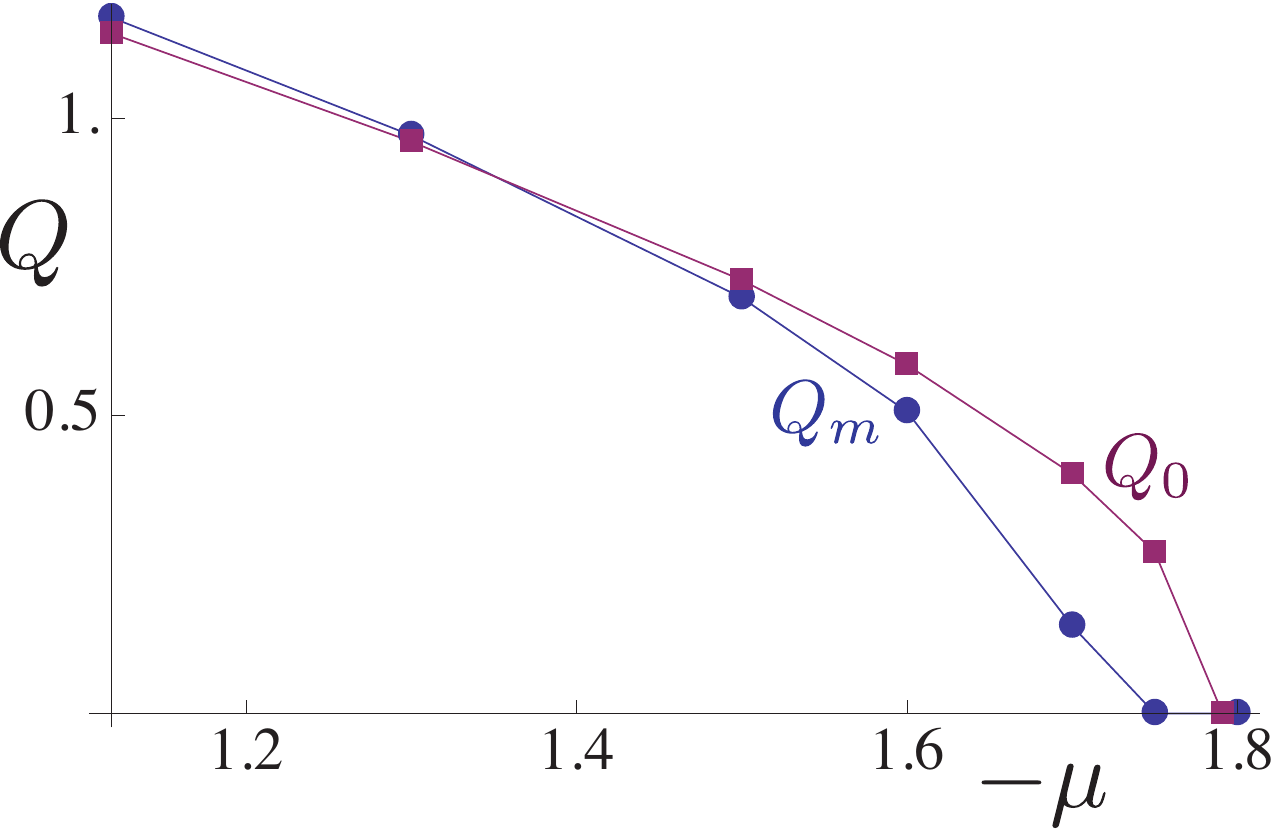}
 \caption{Plot of $Q_m$ (circles), where the minimum of $\lambda_\bQ$ occurs at $\bQ = (\pm Q_m, \pm Q_m)$.
 Also shown are the corresponding values of $Q_0$ (squares), as defined by the hotspots on the Fermi surface in Fig.~\ref{fig:bz}.
 The near equality of $Q_m$ and $Q_0$ is evidence for the pseudospin symmetry; note that this holds even though $\chi_0 (\bq)$ in Eq.~(\ref{chiA}) is 
 peaked at the wavevectors $\bK = (\pi,\pm 3 \pi/4), (\pm 3 \pi/4,\pi)$, as is the case in many hole-doped cuprates.
 }
\label{fig:min}
\end{center}
\end{figure}

The preferred value of $\bQ$ in our Hartree-Fock computation in a metal has the form $\bQ = (\pm Q_m, \pm Q_m)$ \cite{max}; similar
orders have appeared in recent computations \cite{max,metzner,efetov} using the renormalization group
and other methods. 
At low doping, we find that $Q_m \approx Q_0$, where $Q_0$ is defined geometrically from the `hot spots'
on the Fermi surface, as shown in Fig.~\ref{fig:bz} (see also Fig.~\ref{fig:min} for a comparison between the values of $Q_m$ and $Q_0$).
Remarkably, the hot spots of commensurate N\'eel order 
play a crucial role when the antiferromagnetic correlations are short-ranged, and even when they are incommensurate.
Recent field-theoretic studies \cite{max,efetov}
focused on the Fermi surface in the immediate
vicinity of these hot spots, and this connection allows us to interpret our Hartree-Fock results in terms of an emergent approximate pseudospin 
symmetry \cite{max}.
The pseudospin symmetry `rotates' $d$-wave Cooper
pairs to particle-hole pairs: the Cooper pair amplitude $\Delta_S (\bk)$ rotates into $\Delta_\bQ (\bk)$, which explains the predominant
$d$ symmetry of the latter. Our results show that the pseudospin symmetry is a good guide to picking optimum states in lattice
computations on models with short spin correlation lengths, even though the symmetry is exact only in a continuum limit where
the spin correlation length becomes very large.

As in Ref.~\cite{max,efetov}, we propose that the high $T$ pseudogap
is a metal with a fluctuating multi-dimensional order with both a superconducting component, $\Delta_S (\bk)$, and a bond order component
$\Delta_\bQ (\bk )$ over a range of $\bQ$ around $(\pm Q_m, \pm Q_m)$. 
At lower $T$, superconductivity appears by the polarization
of this order along $\Delta_S (\bk)$. Subsequent static charge-ordering requires computation of the non-zero $\bQ$ instabilities
within the superconductor. Fortunately, the latter computation has already been performed in 
closely related models \cite{vojta,vojta2}: bond order modulations were found with $\bQ$ along the $(1,0), (0,1)$ 
directions, as is the case in observations
at low $T$ \cite{jenny,ali,kohsaka,eric,lawler,mesaros,kohsaka2,keimer,chang,hawthorn,hawthorn2}.
In our metallic computations, there is a `valley' of stability from $(Q_m, Q_m)$ to $(Q_m,0)$ and $(0, Q_m)$, but the global minimum is at $(Q_m, Q_m)$ (see Fig.~\ref{fig:lambda}); within the superconducting phase in zero magnetic field, the balance can evidently be tipped
in favor of bond order near $(\pm Q_m , 0)$ and $(0, \pm Q_m)$. The choice of ordering wavevectors between
$(\pm Q_m, \pm Q_m)$ and $(\pm Q_m, 0)$, $(0, \pm Q_m)$ surely depends upon details of the Hamiltonian, the 
value of $T$, and the presence of a magnetic field, 
and is perhaps not accurately estimated by our present simple Hartree-Fock computation. Nevertheless, we expect the 
predominant $d$ symmetry of particle-hole pair condensate, $\Delta_\bQ (\bk)$, to be robust for $|\bQ| \lesssim 2Q_m$, 
for the same reason it is robust for the particle-particle condensate, $\Delta_S (\bk)$, of the superconductor.
 
We examine the following Hamiltonian of electrons on a square lattice of sites $i$ at positions $\br_i$
with annihilation operators $c_{i\alpha}$, where $\alpha = \uparrow, \downarrow$ is a spin label:
\beq
H = \sum_{i,j} \Biggl[ \left(- \mu \delta_{ij} - t_{ij} \right) c_{i \alpha}^\dagger c_{j \alpha}^{\vphantom \dagger}
+ \frac{1}{2} J_{ij} \, \vec{S}_i \cdot \vec{S}_j \Biggr].
\eeq
Here $\mu$ is the chemical potential, $t_{ij}$ are the electron hopping amplitudes, $J_{ij}$ are exchange interactions,
and the electron spin operator $\vec{S}_i = \frac{1}{2} c_{i\alpha}^\dagger \vec{\sigma}_{\alpha\beta} c_{i \beta}^{\vphantom \dagger}$, with $\vec{\sigma}$
the Pauli matrices. The pseudospin symmetry acts as on the Nambu spinor $\Psi_{i \alpha} = ( c_{i \alpha}, \epsilon_{\alpha\beta} c_{i \beta}^\dagger)$ as a SU(2) rotation $V_i$ in Nambu space 
under which $\Psi_{i \alpha}  \rightarrow V_i \Psi_{i \alpha}$. A key property is that $\vec{S}_i = 
\frac{1}{4} \Psi_{i\alpha}^\dagger \vec{\sigma}_{\alpha\beta} \Psi_{i \beta}^{\vphantom \dagger}$,
and this is {\em invariant\/} under the pseudospin transformation. Consequently 
the exchange interaction is invariant under independent rotations $V_i$ on each lattice site \cite{affleck},
and this gauge invariance was
exploited in the study of spin liquid ground states of Mott insulators \cite{leewen}. 
The pseudospin symmetry is completely broken by the $t_{ij}$ terms
in $H$, and so it was expected that pseudospin symmetry plays no role in metallic states, except those that are proximate to certain
spin liquids \cite{kotliar2,leewen}. 
Here, we are interested in metallic states proximate to systems with long-range antiferromagnetism; surprisingly, it was shown in Ref.~\cite{max}, that an analog of the pseudospin gauge 
symmetry of Refs.~\cite{affleck,kee} reappears in the critical theory of the antiferromagnetic quantum critical point in a conventional metal, as 
4 independent global SU(2) pseudospin rotations, one for each pair of hot spots. These rotations serve to map the $d$-wave Cooper pairing
$\Delta_S (\bk)$ to the $d$-wave bond order $\Delta_{\bQ} (\bk)$, as is also evident in our computations below. 

Note that $H$ does not contain an explicit on-site interaction, the `Hubbard $U$'. Both the Cooper pair and the bond order have small on-site components because of the $d$ symmetry, and so $U$ is not important in selecting the ordering instabilities. The effects of $U$ can be accounted for by `slave particle' methods \cite{kotliar,vojta},
and its main consequence is a renormalization of the quasiparticle dispersion.
Finally, such local interactions are irrelevant in the field theory of Ref.~\cite{max}.

For our charge-ordering Hartree-Fock analysis, we need the best variational estimate for the mean-field Hamiltonian
\beq
H_{MF} =  \sum_{i,j}  \left(- \mu \delta_{ij} - t_{ij} - \Delta_{ij} \right) c_{i \alpha}^\dagger c_{j \alpha}^{\vphantom \dagger}
\eeq
where the non-local charge order $\Delta_{ij}$ is written as
\beq
\Delta_{ij} = \sum_{\bQ} \left[ \frac{1}{V} \sum_{\bk} e^{i \bk \cdot (\br_i - \br_j)} \Delta_\bQ (\bk) \right] e^{i \bQ \cdot (\br_i + \br_j)/2} ,
\label{Dij}
\eeq
with $V$ the system volume. This expression highlights the physical interpretation of $\Delta_\bQ (\bk)$:
({\em i\/}) if $\Delta_\bQ (\bk)$ is a constant independent of $\bk$ ({\em i.e.\/} $s$-wave) then we have an ordinary site charge density wave
at wavevector $\bQ$ with only $\Delta_{ii}$ non-zero; ({\em ii\/}) if $\Delta_\bQ (\bk) \sim c_1 \cos k_x + c_2 \cos k_y $
($d$- and extended $s$-wave) then we have bond order at wavevector $\bQ$ with $\Delta_{ij}$ non-zero only if $i$ and $j$ are nearest
neighbors. Also note that hermiticity requires $\Delta_{\bQ}^\ast  (\bk) = \Delta_{-\bQ} ( \bk)$, and under time-reversal $\mathcal{T}: \Delta_{\bQ} (\bk) \rightarrow \Delta_{\bQ} (-\bk)$.

All the functions $\Delta_\bQ (\bk)$ are variational parameters, to be optimized by minimizing
the free energy by
$F \leq F_{MF} + \langle H - H_{MF} \rangle_{MF}$,
where the average is over a thermal ensemble defined by $H_{MF}$. Here, we expand the r.h.s. in 
 powers of $\Delta_\bQ (\bk)$, and replace the inequality by an equality. To quadratic order in $\Delta_{\bQ}$, we write the result in 
terms of hermitian functional operators on the Brillouin zone as
\beq 
F =  \sum_{\bk,\bk',\bQ}\Delta_\bQ^\ast (\bk) \sqrt{\Pi_\bQ (\bk)}
\mathcal{M}_\bQ (\bk, \bk') \sqrt{\Pi_\bQ (\bk')} \Delta_\bQ (\bk') + \ldots \label{FM}
\eeq
where the kernel is
\beq
\mathcal{M}_\bQ (\bk, \bk') =\delta_{\bk ,\bk'} +  \frac{3}{V} \, \chi_0 (\bk - \bk') \sqrt{\Pi_\bQ (\bk) \Pi_{\bQ} (\bk')} \label{kernel}
\eeq
while the polarizability and susceptibility are
\beq
\Pi_\bQ (\bk) =  \frac{f(\varepsilon (\bk + \bQ/2)) - f (\varepsilon (\bk-\bQ/2))}{\varepsilon (\bk-\bQ/2) - \varepsilon (\bk + \bQ/2)}
\quad,\quad \chi_0 (\bq) = \frac{1}{4} \sum_{j} J_{ij} e^{i \bq \cdot (\br_i - \br_j)} , \label{chi0}
\eeq
with $\varepsilon(\bk)$ the electron dispersion associated with $t_{ij}$, and $f$ the Fermi function.
From Eq.~(\ref{FM}) we see that the linear charge-ordering instability of the metal occurs via condensation in the eigenmodes of the operator $\mathcal{M}_{\bQ} (\bk, \bk') $ with the lowest eigenvalues. We have chosen the specific forms of the kernel in Eq.~(\ref{kernel}) so that
we need only solve the following eigenvalue problem 
\beq
\frac{3}{V} \sum_{\bk'} \sqrt{\Pi_{\bQ}(\bk )} \, \chi_0 (\bk - \bk' ) \, \sqrt{\Pi_{\bQ} (\bk' )}\,
\phi_{\bQ} (\bk') = \lambda_{\bQ} \phi_{\bQ} (\bk) \nonumber
\eeq
for the minimum eigenvalues $\lambda_{\bQ}$ and corresponding eigenvectors $\phi_{\bQ} (\bk )$, and their structure is independent of the overall strength of the interaction $\chi_0$. The charge-order will  then be $\Delta_{\bQ} (\bk) \propto \phi_{\bQ} (\bk)/\sqrt{\Pi_{\bQ} (\bk)}$.
Our principal numerical results below are on the $\bQ$ dependence of $\lambda_{\bQ}$, and on the $\bk$ dependence 
of $\Delta_{\bQ} (\bk)$ so obtained. 

We also solved for the corresponding instability of the metal to the superconductor. In this case $H_{MF}$ has the charge-ordering term
$\Delta_{ij}$ replaced by the pairing term $- \sum_{\bk} \Delta_S (\bk) c_{\bk \uparrow} c_{-\bk \downarrow} + \mbox{H.c.}$,
and the subsequent expressions have the replacements $\Delta_\bQ (\bk) \rightarrow \Delta_S (\bk)$, $\mathcal{M}_\bQ (\bk, \bk')
\rightarrow \mathcal{M}_S (\bk, \bk')$, $\Pi_Q (\bk) \rightarrow \Pi_S (\bk)$, $\lambda_\bQ  \rightarrow \lambda_S$, with
$
\Pi_S (\bk) = (1 - 2 f (\varepsilon (\bk)))/(2 \varepsilon (\bk)).$
In particular, the expression for the kernel $\mathcal{M}_S (\bk, \bk')$ in terms of $\Pi_S (\bk)$ has a form identical
to Eq.~(\ref{kernel}), a key consequence of the pseudospin symmetry of the exchange interaction.
Note also that for dispersions with $\varepsilon (\bk + \bQ) = - \varepsilon (\bk)$ we have $\Pi_{\bQ} = \Pi_S$
and so $\mathcal{M}_\bQ = \mathcal{M}_S$;
Ref.~\cite{max} pointed out that the dispersion obeys such a relationship close to the hot spots of a generic Fermi surface
for $\bQ = (\pm Q_0, \pm Q_0)$ (see Fig.~\ref{fig:bz})), and this then establishes the 
pseudospin rotation symmetry between $\Delta_S$ and $\Delta_\bQ (\bk)$.

We assume an electronic dispersion $\varepsilon (\bk ) = - 2 t_1 \left( \cos (k_x) + \cos (k_y) \right) - 4 t_2 \cos (k_x) \cos (k_y)
- 2 t_3 \left( \cos (2 k_x) + \cos (2 k_y) \right) - \mu$ and a susceptibility $\chi_0 (\bq)$ which is peaked 
near the antiferromagnetic wavevector
\beq
\chi_0 (\bq) = \sum_{\bK} \frac{A}{4(\xi^{-2} + 2 ( 2 - \cos (q_x - K_x) - \cos (q_y - K_y) ))}, \label{chiA}
\eeq
where $\xi$ is the antiferromagnetic correlation length, the sum extends over 
$\bK = \pm (\pi, \pi ( 1 - \delta)),~\pm (\pi(1-\delta),\pi)$, and we used both the commensurate case $\delta =0$
and the incommensurate case $\delta = 1/4$, with little difference between the results. We only need a short spin correlation length, $\xi$, 
and indeed obtained very similar results even for the case where $\chi_0 (\bq)$ was obtained from Eq.~(\ref{chi0}) with only a nearest-neighbor $J_{ij}$.
We diagonalized the kernels after discretizing the Brillouin zone to $L^2$ points 
with $L$ up to 80, and the results below are for 
$t_1 =1$, $t_2 = -0.32$, and $t_3=0.128$ for a range of values of $T$, $\mu$,  and $\xi$.

{\textit{Numerical results}}.
For the full range of parameters examined, we consistently found that $\lambda_S$ was the minimal eigenvalue
(indeed, BCS theory implies $-\lambda_S$ diverges logarithmically as $T \rightarrow 0$), 
and the corresponding eigenvector $\Delta_S ( \bk )$ was well approximated 
by the $d$-wave form $\sim (\cos k_x - \cos k_y)$ (see Table~\ref{table}). 
So $d$-wave superconductivity is the primary instability.

For the charge ordering instabilities, we show 
the $\bQ$ dependence of
$\lambda_\bQ$ in Figs~\ref{fig:lambda} and in the supplement.
We characterize the corresponding eigenvectors $\Delta_{S,\bQ} (\bk)$ using
orthonormal basis functions, $\psi_\gamma (\bk)$ of the square
lattice space group:
\beq
\Delta_\bQ (\bk) = \sum_\gamma c_{\bQ,\gamma} \, \psi_\gamma (\bk) \label{expand}
\eeq
where $c_{\bQ,\gamma}$ are numerical coefficients collected in Table~\ref{table}.
\begin{table}
\centerline{
\begin{tabular}{|c|c|c|c|c|c|c|}
\hline 
$\gamma$ & $\psi_\gamma (\bk)$ & $\bQ = $ & $\bQ = $ &  $\bQ = $ &  $\bQ = $ & $\Delta_S (\bk)$ \\
& & $(Q_m,Q_m)$ & $(Q_m, 0)$ & $(0,0)$&$(\pi,\pi)$ & \\
\hline
$s$ & 1 & 0 & -0.226 & 0 & 0 & 0\\
$s'$ & $\cos k_x + \cos k_y$ & 0 &  0.040 & 0 & 0 & 0 \\
$s''$ & $\cos (2k_x) + \cos (2k_y)$ & 0 & -0.051 & 0 & 0 & 0  \\
$d$ & $\cos k_x - \cos k_y$ & 0.993 & 0.964 & 0.997 & 0 & 0.998 \\
$d'$ & $\cos (2k_x) - \cos (2k_y)$ & - 0.058 & -0.057 & -0.044 & 0 &  -0.047 \\
$d_{xy}$ & $2 \sin k_x \sin k_y$ & 0 &  0 & 0 & 0 & 0\\
$p_x$ & $\sqrt{2} \sin k_x$ & 0 & 0 & 0 & 0.706 & 0 \\
$p_y$ & $\sqrt{2} \sin k_y$ & 0 & 0 & 0 & -0.706 & 0 \\
$g$ & $(\cos k_x - \cos k_y) $ & -0.010 & 0 & 0 & 0 & 0 \\
 & $\times \sqrt{8} \sin k_x \sin k_y $ &  &  & & & \\
\hline
\end{tabular}
}
\vspace{0.1in}
\caption{Values of $c_{\bQ,\gamma}$ in the expansion for $\Delta_\bQ (\bk)$ in  Eq.~(\ref{expand}) for various values $\bQ$ and $\gamma$.
The values of $c_{\bQ,\gamma}$ are normalized so that $\sum_\gamma |c_{\bQ,\gamma}|^2 = 1$, where the sum over $\gamma$ includes the small
contributions from higher order basis functions not shown above. Values shown as 0
are constrained to be exactly zero by symmetry.
The last column shows the coefficients in the corresponding expansion for $\Delta_S (\bk)$. Parameters are as in Fig.~\ref{fig:lambda}, and $Q_m = 4 \pi/11$.}
\label{table}
\end{table}
Depending upon the symmetry of $\bQ$ (in particular, the little group of the wavevector $\bQ$) and of the eigenvector,
some of the $c_{\bQ,\gamma}$ may be exactly zero. But for a generic $\bQ$, only time-reversal constrains the values of $c_{\bQ, \gamma}$,
and we are allowed to have an admixture of many basis functions. Nevertheless, only a small number of basis functions
have appreciable coefficients, and so Eq.~(\ref{expand}) represents a useful expansion.

The global minimum of $\lambda_\bQ$ is at a wavevector along the diagonal with $\bQ = (Q_m,Q_m)$, and we show a plot of $Q_m$ as
a function of chemical potential in Fig.~\ref{fig:min}.
We also show the corresponding values of $Q_0$ as defined in Fig.~\ref{fig:bz}; for small doping we see that $Q_m \approx Q_0$, one of our key results: the pseudospin symmetry of the hot-spot theory of Ref.~\cite{max} is a good guide to determining the ordering even for models with short-range, incommensurate, antiferromagnetic spin correlations.
 At larger doping, after the chemical potential crosses the van-Hove singularity \cite{metzner}, there are no hot spots, and we find $Q_m = 0$. For $\bQ = (Q_m, Q_m)$, Table~\ref{table} shows that
 $\Delta_\bQ (\bk)$ is predominantly $d$, with a small
admixture of $g$. 
For $\bQ = (Q_m,0)$, $\Delta_\bQ$ remains predominantly $d$, but now has a small $s$ component \cite{vojta2}.
 
At $\bQ=0$, we find that $\Delta_\bQ (\bk)$ is purely $d$: this corresponds to Ising-nematic order \cite{KFE98,YK00,HM00}. The $\mathcal{T}$-breaking 
`circulating-current' order of Ref.~\cite{varma} has a $p_{x,y}$ form for $\Delta_\bQ (k)$ at $\bQ = 0$,
but this does not appear as a lowest eigenvalue, and so is not present in Fig.~\ref{fig:lambda}.
Finally, $\lambda_\bQ$ also has a broad local minimum at $\bQ = (\pi, \pi)$: here
$\Delta_\bQ (\bk)$ does have the $p_{x,y}$ form which breaks $\mathcal{T}$, and leads to the 
state with spontaneous orbital currents \cite{kotliar,sudip,leewen}.

Experiments \cite{jenny,ali,kohsaka,eric,lawler,mesaros,kohsaka2,keimer,chang,hawthorn,hawthorn2} 
have observed charge ordering at $\bQ = (Q_m,0), (0, Q_m)$
at low $T$. Choosing the largest 2 components at this wavevector from Table~\ref{table}, we have 
\beq
 \Delta_\bQ (\bk) = \left\{ \begin{array}{ccc} \Delta_s + 
\Delta_d ( \cos k_x - \cos k_y ) &~,~&\bQ = (\pm Q_m,0) \\
 \Delta_s - 
\Delta_d ( \cos k_x - \cos k_y ) &~,~&\bQ = (0,\pm Q_m) \end{array} \right. 
\label{Deltasd}
\eeq
with $\Delta_s/\Delta_d = -0.234$. Similarly, we can have bond-ordering along $\bQ = (\pm Q_m, \pm Q_m)$ with only $\Delta_d$ non-zero.
We present implications of these orders
for X-ray scattering, nuclear magnetic resonance, photoemission
and scanning tunneling microscopy in the supplement.

Our evidence for pseudospin symmetry between Cooper pairing and charge order should have significant implications for 
the dynamics of these orders, which have been studied recently in Ref.~\cite{joeoren}.
For the phase diagram of the hole-doped cuprates, our model has a $T=0$ quantum-critical point near optimal doping associated with
disappearance of this bond order \cite{vojta,julien}.
 An important challenge is to use such a critical
point to describe the evolution of the Fermi surface \cite{suchitra2}, and  the `strange' metal.

{\textit{Acknowledgments}}. 
We thank for A.~Chubukov, D. Chowdhury, J.~C.~Davis, E.~Demler, K.~Efetov, D.~Hawthorn, P.~Hirschfeld,
J.~Hoffman, M.-H. Julien, 
E.-A.~Kim, S.~Kivelson, 
G.~Kotliar, M.-H.~Julien, H.~Meier, W.~Metzner, C.~P\'epin, C.~Proust, S.~Sebastian, L. Taillefer,  and M.~Vojta for  
useful discussions. This research was supported by the NSF under Grant DMR-1103860, the U.S. Army Research
Office Award W911NF-12-1-0227, and the John Templeton Foundation.

\newpage
\section{Supplementary material}

\begin{figure}[h!]
\begin{center}
 \includegraphics[width=3.8in]{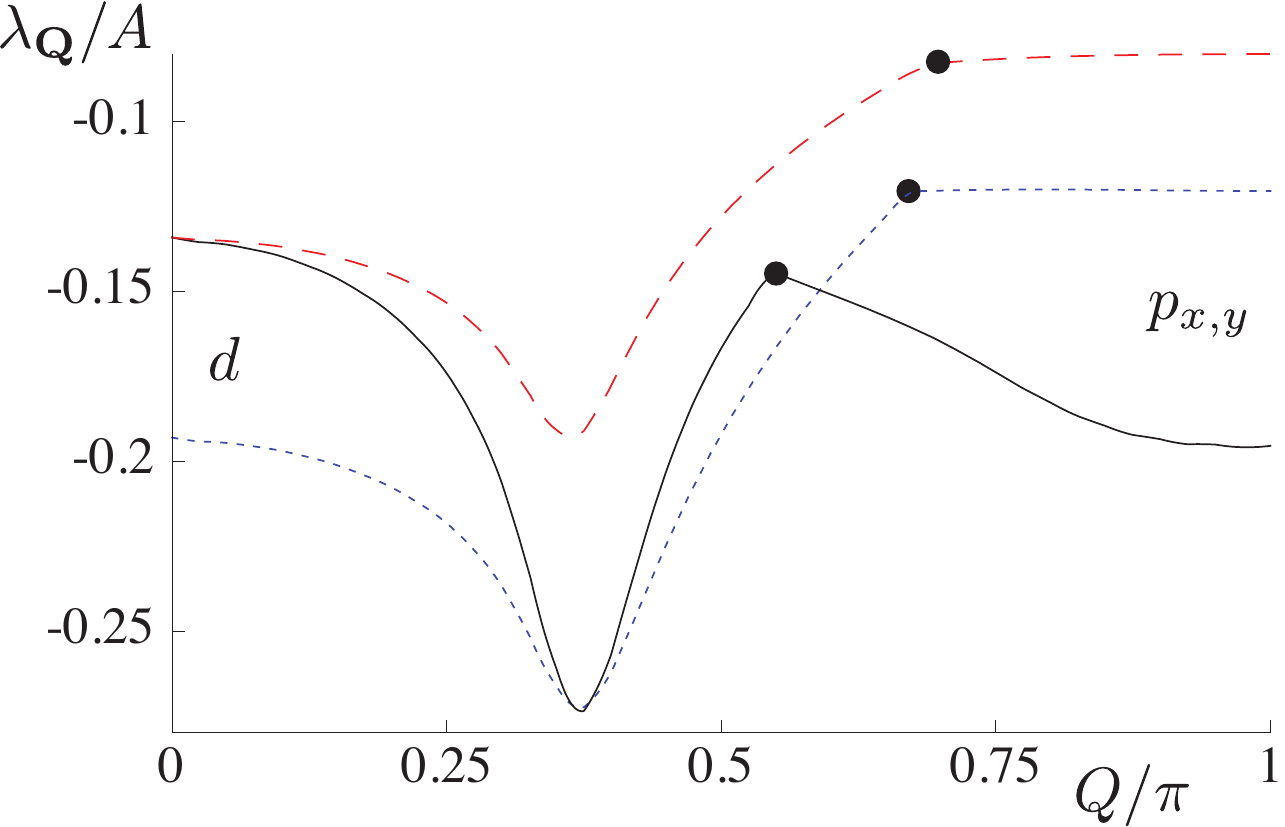}
 \caption{Plot of the eigenvalue of Fig.~\ref{fig:lambda} along the Brillouin zone diagonal
 with $\bQ = (Q, Q)$ (full line), along the line $\bQ = (Q_m,Q)$ (dotted blue line),
 now with $L=80$.
The eigenfunction $\Delta_\bQ (\bk )$ has
 predominant $d$ symmetry 
 (as in the state of Ref.~\cite{max}) with $\mathcal{T}$ preserved to the left of the filled circles, and predominant $p_{x,y}$ symmetry with $\mathcal{T}$ broken (as in the state of Refs.~\cite{kotliar,sudip}) to the right of the filled circles. The $\bQ = (0,0)$ point corresponds
 to Ising nematic order \cite{KFE98,YK00,HM00}.
 }
\label{fig:lambda1d}
\end{center}
\end{figure}

First, we give further details on the function $\lambda_\bQ$ in Fig.~\ref{fig:lambda}.
In Fig.~\ref{fig:lambda1d}, we plot $\lambda_\bQ$ along different lines in the Brillouin zone, and also
indicate the regions where $\mathcal{T}$ is preserved and broken.
\begin{figure}
\begin{center}
 \includegraphics[width=3.8in]{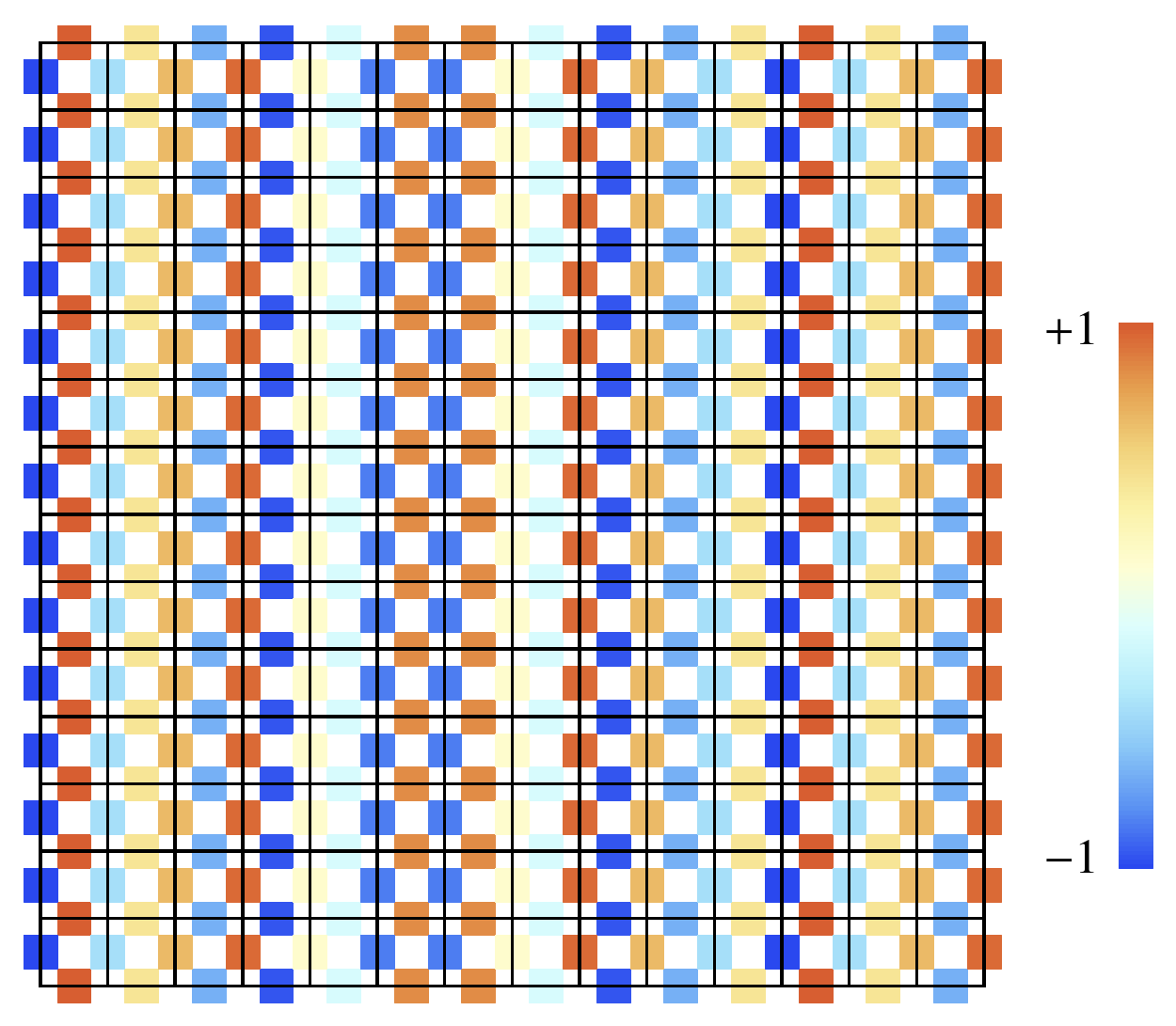}
 \caption{Plot of the values of $\Delta_{ij}$, when $i$ and $j$ are nearest neighbors; the value is denoted by a colored square
 centered at the midpoint between $i$ and $j$. The lines intersect at the Cu sites, and the colored squares are on the O sites: the colors are therefore a measure of the charge
 density (or other spectral properties) on the O sites. This is also
 the bond-component of the ordering in Eq.~(\ref{Deltasd}), proportional to $\Delta_d$; 
 there is an additional site-component, proportional to $\Delta_s$, which is 
 not shown. The plot above is 
 for the case of uni-directional order at $\bQ = (\pm Q_m, 0)$ where $Q_m = 4 \pi/11$, 
 and other cases are in the following figures.
 }
\label{fig:qdw1}
\end{center}
\end{figure}

\begin{figure}
\begin{center}
 \includegraphics[width=3.8in]{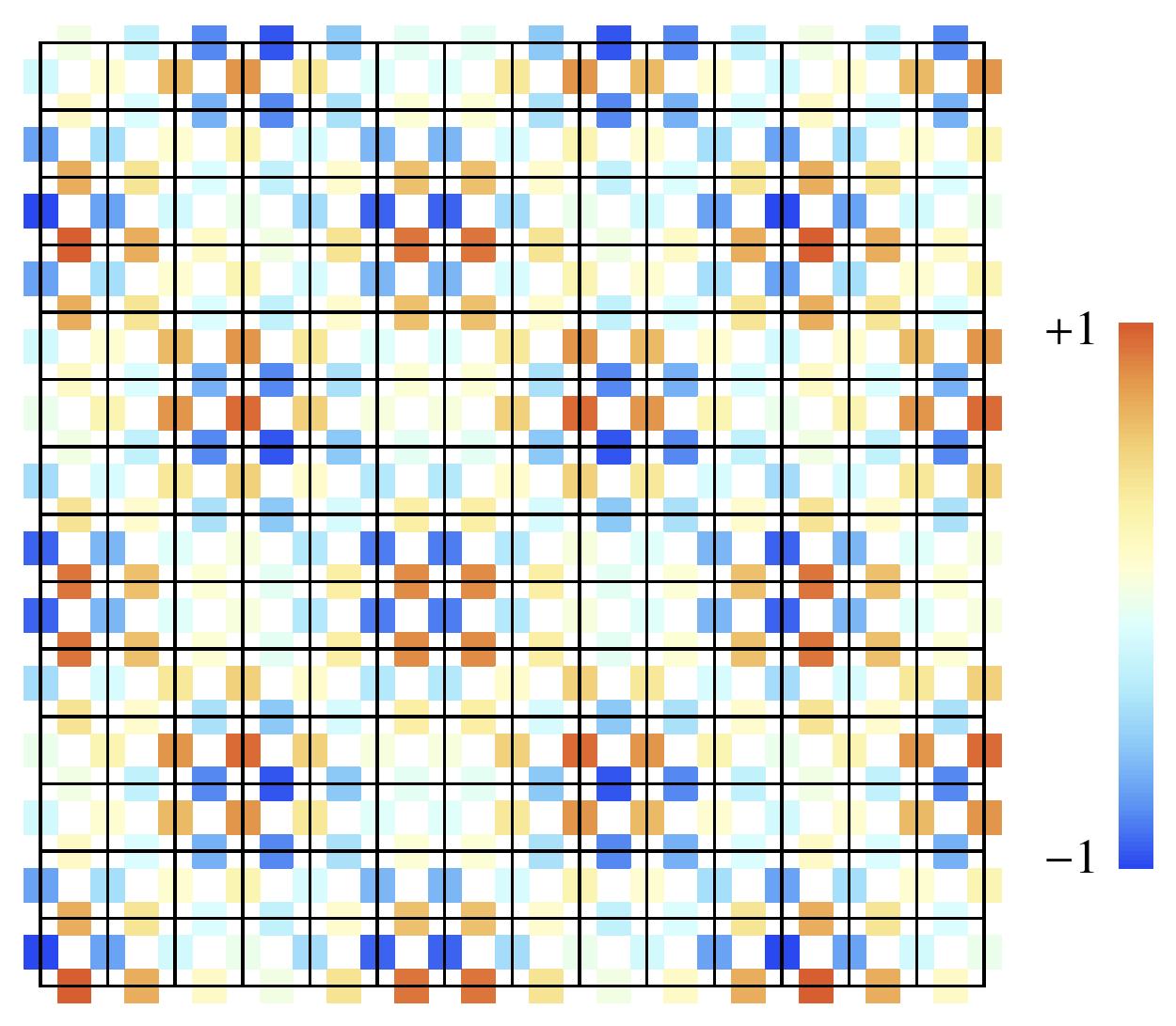}
 \caption{As in Fig.~\ref{fig:qdw1}, but for the case of bi-directional order at $\bQ = (\pm Q_m, 0)$ and 
$\bQ = (0,\pm Q_m)$.}
\label{fig:qdw2}
\end{center}
\end{figure}

Next, we describe properties of the bond-ordered state in Eq.~(\ref{Deltasd}). 
Inserting Eq.~(\ref{Deltasd}) into Eq.~(\ref{Dij}), we see that the real space order parameter $\Delta_{ij}$ is non-zero 
only when $i=j$, or when $i$ and $j$ are nearest neighbors. The values of $\Delta_{ii}$ correspond to an ordinary on-site charge
density wave on the Cu sites at wavevectors $\bQ = (0, \pm Q_m), (\pm Q_m, 0)$ with amplitude proportional to $\Delta_s$.
The larger component of the ordering is however the bond-density wave given by $\Delta_{ij}$ with $i$,$j$ nearest neighbors,
whose amplitude is proportional to $\Delta_d$. We show plots of the values of $\Delta_{ij}$ on the bonds of the square lattice
in Fig.~\ref{fig:qdw1} and~\ref{fig:qdw2}. 
Fig~\ref{fig:qdw1} contains the case of uni-directional order only at the wavevectors 
$\bQ = (\pm Q_m, 0)$, while Fig.~\ref{fig:qdw2} is the case of bi-directional order at wavevectors $\bQ = (\pm Q_m, 0)$ and 
$\bQ = (0,\pm Q_m)$. 

\begin{figure}
\begin{center}
 \includegraphics[width=3.8in]{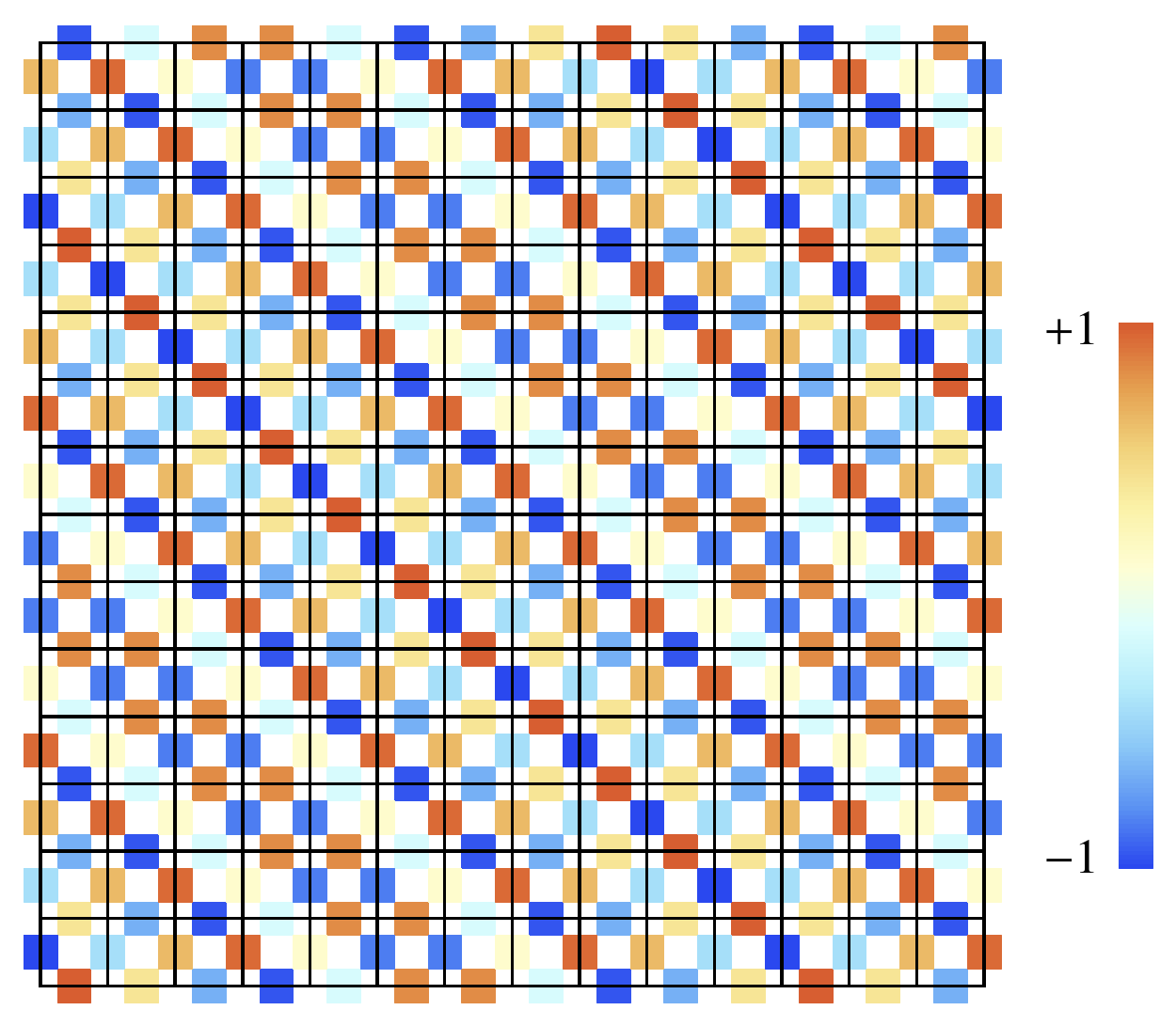}
 \caption{As in Fig.~\ref{fig:qdw1}, but for the case of uni-directional order at $\bQ = \pm(Q_m, Q_m)$. 
 We have chosen $\Delta_\bQ (\bk)$ to be purely $d$, which is an excellent approximation to the state in Table~\ref{table}.
 In this case $\Delta_{ij}$ is non-zero only if $i$ and $j$ are nearest neighbors, and these are shown above; there is no density wave
 on the Cu sites. This plot also appeared in Ref.~\cite{max} with a different period.
 }
\label{fig:qdw3}
\end{center}
\end{figure}
\begin{figure}
\begin{center}
 \includegraphics[width=3.8in]{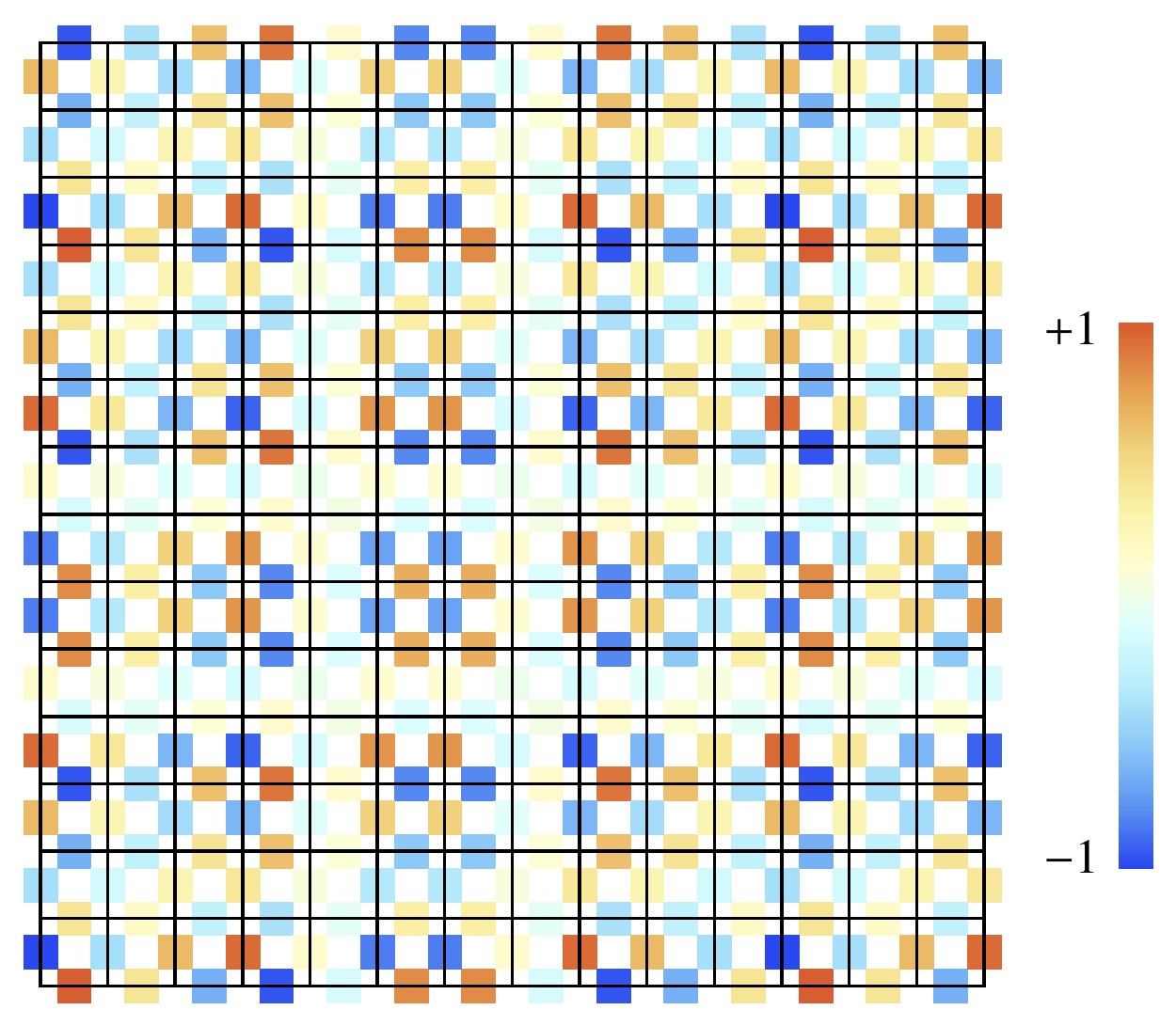}
 \caption{As in Fig.~\ref{fig:qdw3}, but for the case of bi-directional order at $\bQ = \pm( Q_m, Q_m)$ and 
$\bQ = \pm (Q_m, -Q_m)$.}
\label{fig:qdw4}
\end{center}
\end{figure}
For completeness, we also show the corresponding plots for ordering along
 $\bQ = (\pm Q_m, \pm Q_m)$ in Figs~\ref{fig:qdw3} and~\ref{fig:qdw4}; 
these appeared earlier
in Figs.~22 and 23 in Ref.~\cite{max} at a different period. 
Note that the difference between bi-directional order
at $\bQ = (\pm Q_m, 0)$ and 
$\bQ = (0,\pm Q_m)$ in Fig.~\ref{fig:qdw2} and bi-directional order at $\bQ = \pm( Q_m, Q_m)$ and 
$\bQ = \pm(Q_m, - Q_m)$ in Fig.~\ref{fig:qdw4} is subtle, 
and not immediately apparent at first glance: the periods along the $x$ and $y$ axes appear the same.
However, the Fourier transforms of these two cases are distinct.

The four plots in Fig.~\ref{fig:qdw1}-\ref{fig:qdw4} 
together contain information that should be useful in interpreting scanning tunneling microscopy, nuclear magnetic resonance, and X-ray scattering experiments: 
the colors can be viewed as a measure of any observable on the O site which is invariant under time-reversal
and spin rotation. Most simply, such an observable is the charge density on the O site, but any spectral property of the O atom
also qualifies, and the latter can have readily measurable consequences in such experiments.

\begin{figure}[h]
\begin{center}
 \includegraphics[width=5.5in]{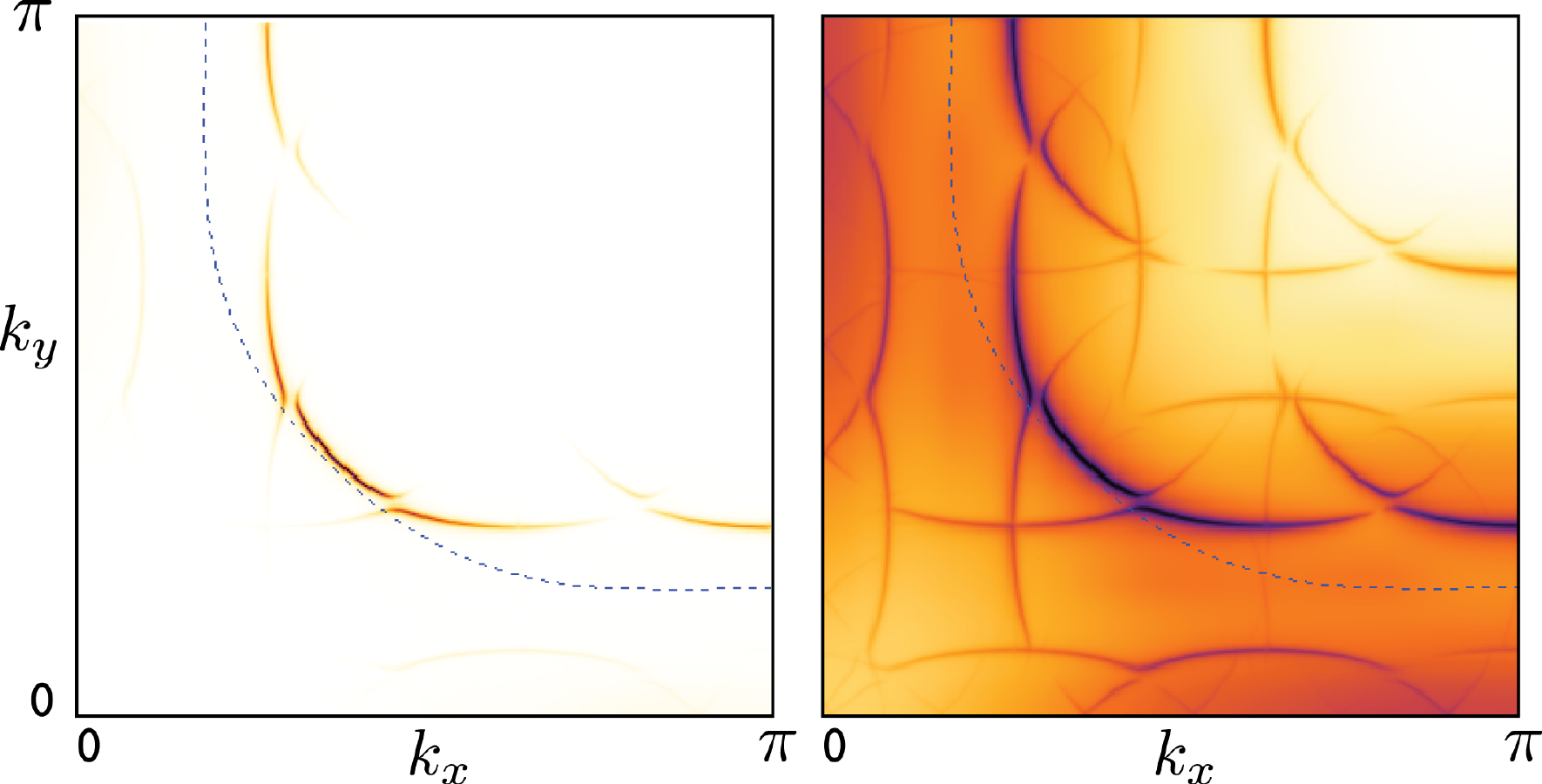}
 \caption{Electron spectral density in the phase with bidirectional charge order at $\bQ = (Q_m,0)$
 and $(0, Q_m)$ with $Q_m=4 \pi/11$. The left panel show $\mbox{Im} G_{\bk, \bk} (\omega + i \eta)$
 at $\omega = 0$ and $\eta = 0.02$; the right panel shows $\log \left[\mbox{Im} G_{\bk, \bk} (\omega + i \eta) \right]$
 for the same parameters, as a way of enhancing the low intensities. The dashed line is the underlying Fermi surface of the metal
 without charge order.
  The charge order is as in Eqs.~(\ref{Dij},\ref{Deltasd}) with $\Delta_d = 0.3$, $\Delta_s/\Delta_d = -0.234$, and other parameters as in Fig.~\ref{fig:lambda}.}
\label{fig:spec}
\end{center}
\end{figure}

\begin{figure}[h]
\begin{center}
 \includegraphics[width=5.5in]{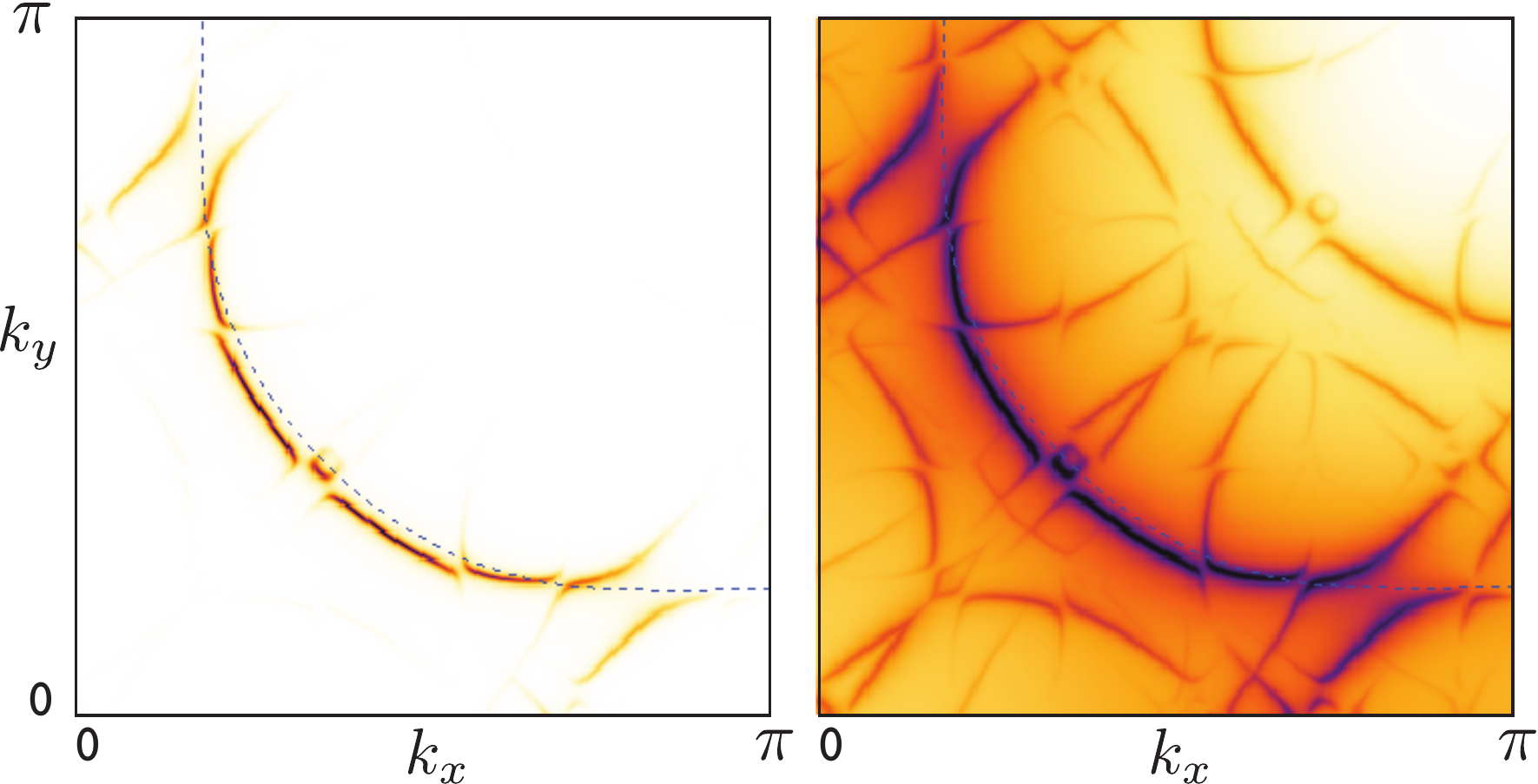}
 \caption{As in Fig.~\ref{fig:spec}, but for the case of bi-directional ordering along $\bQ = \pm (Q_m, Q_m),  \pm(Q_m, -Q_m)$.
  The charge order is as in Eqs.~(\ref{Dij},\ref{Deltasd}) with $\Delta_d = 0.3$, and $\Delta_s=0$ is required by symmetry.}
\label{fig:spec2}
\end{center}
\end{figure}

Finally, we consider the electronic spectral function in the presence of bond-ordering in a metal. This is obtained by
diagonalizing the following Hamiltonian
\beq
H_b = \sum_\bk \Biggl[ \varepsilon (\bk) c_{\bk \alpha}^\dagger c_{\bk \alpha}^{\vphantom \dagger}
- \sum_{\bQ} \Delta_\bQ ( \bk + \bQ/2) \, c_{\bk + \bQ, \alpha}^\dagger c_{\bk\alpha}^{\vphantom\dagger} \Biggr],
\label{Hb}
\eeq
where the sum over $\bk$ extends over the complete Brillouin zone of the square lattice.
For the case of bi-directional order 
in Eq.~(\ref{Deltasd}), the sum over $\bQ$ extends
over the 4 values $(\pm Q_m, 0)$ and 
$(0,\pm Q_m)$. Some care must be taken in evaluating the wavevector $\bQ/2$ in the argument of $\Delta_\bQ$ in Eq.~(\ref{Hb}) as it is not invariant under
translation of $\bQ$ by a reciprocal lattice vector of the square lattice: in each term, we take the momenta $\bk$ and $\bk +\bQ$ to  be separated by exactly $\bQ$ (and not modulo a reciprocal lattice vector), and then $\Delta_{\bQ} (\bk + \bQ/2)$ is evaluated at the midpoint between them.
For $Q_m = 4 \pi/11$, determining the spectrum of $H_b$ involves diagonalizing a $121 \times 121$ matrix for each $\bk$. From the eigenfunctions and eigenvectors we computed the imaginary part of the single-electron Green's function, $\mbox{Im} G_{\bk, \bk} (\omega + i \eta)$,
the quantity related to the photoemission spectrum.
For the bi-directional ordering along $\bQ = (\pm Q_m, 0), (0, \pm Q_m)$ of Eq.~(\ref{Deltasd})
the result is 
shown in Fig.~\ref{fig:spec}.
The corresponding result for bi-directional ordering along $\bQ = \pm (Q_m, Q_m),  \pm(Q_m, -Q_m)$
is in Fig.~\ref{fig:spec2}; in this case $\Delta_s=0$ by symmetry, and only $\Delta_d$ was non-zero. 

The stability of the Fermi arc in the `nodal' region ($k_x \approx k_y$) is enhanced \cite{mohit,vojta2} because of  the weak coupling to the charge order, arising from the predominant
$d$ character of Eq.~(\ref{Deltasd}). In the anti-nodal region, the parent Fermi surface has been gapped out by the bond order,
but `shadows' are apparent at wavevectors shifted by the charge order.
However, these Fermi surfaces 
should be easily broadened by impurity-induced phase-shifts in the charge ordering, while protecting the nodal arcs.
Furthermore, contributions from the superconducting component of the pairing order parameter should also help fully gap out the antinodal
region.

\end{document}